\documentclass{aa}
\usepackage{natbib}
\usepackage{rotating}
\bibpunct{(}{)}{;}{a}{}{,}
\usepackage{graphicx}
\usepackage{txfonts}
\begin{document}
   \title{XMM-{\it Newton} X-ray spectroscopy of the high-mass X-ray
     binary 4U1700$-$37 at low flux}
   \titlerunning{XMM-{\it Newton} X-ray spectroscopy of the high-mass X-ray
     binary 4U1700$-$37 at low flux}
   \author{
             A. van der Meer\inst{1},
             L. Kaper\inst{1},
             T. Di Salvo\inst{1},
             M. M\'endez\inst{2},
	     M. van der Klis\inst{1},
	     P. Barr\inst{3},
             \and
	     N. R. Trams\inst{3}
          }
   \authorrunning{A. van der Meer et al.}
   \offprints{A. van der Meer, \email{ameer@science.uva.nl}}
   \institute{
                Sterrenkundig Instituut ``Anton Pannekoek'' and Center
                for High-energy Astrophysics (CHEAF),
                University of Amsterdam, Kruislaan 403, NL-1098 SJ
                Amsterdam, The Netherlands
                \and
		Space Research Organization of the Netherlands,
                National Institute for Space Research, Sorbonnelaan 2,
                NL-3584 CA Utrecht, The Netherlands
                \and
		Integral Science Operations Centre, Astrophysics Div.,
		Science Department, ESTEC, P.O.Box 299, 
		2200 AG Noordwijk, The Netherlands
             }
   \date{Received ??, ??; accepted ??, ??}

%
%

   \abstract{
  We present results of a monitoring campaign of the high-mass X-ray
  binary system \object{4U~1700$-$37}/\object{HD~153919}, carried out
  with XMM-{\it Newton} in February 2001. The system was observed at
  four orbital phase intervals, covering 37\% of one 3.41-day orbit. The
  lightcurve includes strong flares, commonly observed in this source. We
  focus on three epochs in which the data are not affected by photon pile
  up: the eclipse, the eclipse egress and a low-flux interval in the
  lightcurve around orbital phase $\phi \sim 0.25$. The
  high-energy part of the continuum is modelled as a direct plus a
  scattered component, each represented by a power law with identical
  photon index ($\alpha \sim 1.4$), but with different absorption
  columns. We show that during the low-flux interval the continuum
  is strongly reduced, probably due to a reduction of the accretion
  rate onto the compact object. A soft excess is detected in all
  spectra, consistent with either another continuum component
  originating in the outskirts of the system or a blend of
  emission lines.
  Many fluorescence emission lines from near-neutral species and
  discrete recombination lines from He- and H-like species are
  detected during eclipse and egress. The fluorescence Fe K$\alpha$
  line at 6.4~keV is very prominent; a second K$\alpha$ line is
  detected at slightly higher energies (up to 6.7~keV) and a K$\beta$
  line at 7.1~keV. In the low-flux interval the Fe K$\alpha$ line
  at 6.4~keV is strongly (factor $\sim 30$) reduced in strength. In eclipse, the Fe
  K$\beta$/K$\alpha$ ratio is consistent with a value of 0.13. In
  egress we initially measure a higher ratio, which can be explained
  by a shift in energy of the Fe K-edge to $\sim 7.15$~keV, which is
  consistent with moderately ionised iron, rather than neutral iron,
  as expected for the stellar wind medium. The detection of
  recombination lines during eclipse indicates the presence of an
  extended ionised region surrounding the compact object. The
  observed increase in strength of some emission lines corresponding
  to higher values of the ionisation parameter $\xi$ further
  substantiates this conclusion.

   }

   \maketitle

%
%

\section{Introduction}
\label{intro}
  \object{4U~1700$-$37}/\object{HD~153919} is a high-mass X-ray binary
  (HMXB) discovered by the {\it Uhuru} satellite \citep{jon73}. HMXBs
  are relatively young ($< 10^7$~year) systems composed of a massive
  OB-type star and a compact object, either a neutron star or a black
  hole (for a recent catalogue of X-ray binaries see
  \citealt{liu00}). X-rays are produced when matter from the OB star
  accretes onto the compact object. The majority ($\sim 80\%$) of the
  HMXBs are Be/X-ray binaries. These systems are often X-ray
  transients, only producing X-rays when the compact object accretes
  matter by passing through the dense equatorial disc of a
  Be-star. The other group of systems host a more massive OB
  supergiant (for a review see \citealt{kap01}). About a dozen OB
  supergiant X-ray binary systems are known in the Milky Way and the
  Magellanic Clouds. Among them are a few that host a black-hole
  candidate (e.g. \object{Cyg~X$-$1}).

  Accretion takes place through either the strong stellar wind of the
  optical companion or Roche-lobe overflow. In a wind-fed system the
  X-rays ionise a relatively small region in the stellar wind due to
  the low X-ray luminosity (L$_{\rm X}$ $\sim 10^{35} - 10^{36}$ erg
  s$^{-1}$). The size of the ionisation zone can be determined from
  the orbital modulation of UV resonance lines formed in the wind
  \citep{hat77,loo01}.
  In a Roche-lobe overflow system, matter flows via the inner
  Lagrangian point to an accretion disc. A powerful X-ray source
  (L$_{\rm X}$ $\sim 10^{38}$ erg s$^{-1}$) is produced, which ionises
  almost the whole stellar wind, such that only in the X-ray shadow of
  the optical companion a stellar wind can emerge
  (e.g. \object{LMC~X$-$4}; \citealt{bor99}). For a more general review on
  X-ray binaries we refer to e.g. \citet{nag89} or \citet{lew95}.

  X-ray spectra of HMXBs are often described by a power law with
  photon index $\alpha \sim 1$, modified at high energies (above
  10--20~keV) by an exponential cutoff. A spectrum of this form can be
  produced by inverse Compton scattering of soft X-rays by hot
  electrons in the accretion column near the compact object
  \citep[e.g.][]{hay85}. This continuum component is often scattered by
  the stellar wind of the optical companion. This results in a similar
  component, but with a different absorption column depending on the
  orbital phase of the system.
  In the high energy part ($\gtrsim 10$~keV) of the spectrum of
  most X-ray pulsars broad absorption lines have been discovered,
  which are attributed to cyclotron resonant scattering. From the
  energy of these lines strong magnetic fields are inferred ($B
  \sim 10^{12}$ Gauss; e.g. \citealt{tru78}).

  In some HMXBs a soft excess at $\sim 0.1-1$ keV is detected of which
  the physical origin is not well understood. If this
  component has a thermal origin, its relatively low temperature
  requires in some cases a size of the emission region much larger
  than the size of the compact object, of the order of the
  magnetospheric radius, where the pressure of the infalling matter
  balances the pressure of the magnetic field of the compact object
  (e.g. \object{Cen~X$-$3}; \citealt{bur00}). \citet{hic04} show that a
  soft excess is very common in X-ray binary pulsars and that
  observations that do not show evidence of a soft excess are probably
  limited by high absorption, low flux or insufficient soft
  sensitivity. \citet{hic04} find that the soft excess in luminous
  (L$_{\rm X}$ $\gtrsim 10^{38}$ erg s$^{-1}$) sources can be
  explained by repocessing of hard X-rays from the neutron star by
  optically thick, accreting material. For less luminous (L$_{\rm
  X}$ $\lesssim 10^{36}$ erg s$^{-1}$) sources the soft excess is
  probably due to other processes, e.g. emission from photoionised or
  collisionally heated diffuse gas or thermal emission from the
  surface of the neutron star.

  With the advent of the X-ray observatories {\it Chandra} and
  XMM-{\it Newton} it has become possible to study the X-ray spectra
  of HMXBs in unprecedented detail. Their eclipse spectra show many
  emission lines and radiative recombination continua (e.g. \object{Vela~X$-$1}
  \citealt{sch02,gol04}). Such an emission line spectrum originates in the
  ionisation zone surrounding the X-ray source and becomes detectable
  during eclipse due to the supression of the strong continuum emission
  produced by the X-ray pulsar.
  \citet{pae00} show that the {\it Chandra} spectrum of \object{Cyg~X$-$3} is
  consistent with a cool, optically thin, photoionised plasma in
  equilibrium, whereas the spectrum of \object{4U~1700$-$37}
  shows a more ``hybrid'' plasma in which also evidence is found for a
  collisionally ionised plasma \citep{bor03}. \citet{woj03}
  report on an apparent change of plasma properties from eclipse to
  eclipse egress of \object{Cen~X$-$3}, which they can explain with a
  photoionised plasma for which the effects of resonant line
  scattering are taken into account as well. \citet{wat03} report on a
  fully resolved Compton-scattered Fe~K$\alpha$ fluorescence line in
  \object{GX~301$-$2}. They show that the flux ratio of the line and shoulder
  together with the energy distribution of the shoulder can be used to
  directly constrain the absorption column, temperature and abundances
  of the scattering medium.

\subsection{4U~1700$-$37}
  Located at a distance of 1.9~kpc \citep{ank01}
  \object{4U~1700$-$37} is powered by the dense stellar wind (\.{M}$\sim
  10^{-5}$~M$_{\odot}$~yr$^{-1}$; \citealt{cla02}) of the O6.5~Iaf+
  supergiant \object{HD~153919} \citep{jon73,mas76}. It is the hottest and most
  luminous optical companion known in HMXBs.
  \citet{ank01} propose that the system escaped the OB association Sco OB1
  $\sim$ 2 million years ago, travelling with a velocity of
  $\sim$ 75~km~s$^{-1}$. They show that the progenitor of the compact
  object should have evolved into a supernova within $\sim$ 6 million years (if
  it was born as a member of the association). This corresponds to a
  lower limit of its initial mass of $\sim$ 30 M$_{\sun}$
  \citep{sch92}.

  Periodic X-ray pulsations have not been detected so far (although
  \citealt{bor03} report quasi-periodic oscillations), so that the
  nature of the compact object remains unclear \citep{got86,cla02}. Most
  likely, \object{4U~1700$-$37} is a neutron star since its X-ray spectrum is well
  described by the standard accreting pulsar model \citep{whi83}.
  There are indications that the mass of the compact object is larger
  than 2 M$_{\sun}$ \citep{rub96,cla02}. If true, this might qualify
  it as a low-mass black-hole \citep{bro96} or a massive
  neutron star, like in \object{Vela~X$-$1} \citep{bar01,qua03}.
  \citet{rey99} report the possible presence of a cyclotron feature at
  $\sim 37$~keV in {\it Beppo}SAX observations. If real, \object{4U~1700$-$37}
  has a magnetic field $B \sim 2.3 \times 10^{12}$ Gauss,
  characteristic of a HMXB accreting pulsar.

  This system is unique in showing broad and variable emission lines
  in the ultraviolet part of the spectrum. \citet{kap90} propose that
  these lines are produced by photons from strong EUV emission lines
  close to the \ion{He}{ii} Lyman-$\beta$ transition (256 \AA), which
  are Raman scattered by He~{\sc ii} ions in the stellar wind of the
  optical companion via the corresponding downwards transition near
  the \ion{He}{ii} Balmer-$\alpha$ line at 1640 \AA.
  The input EUV emission lines can not be studied directly due to the
  opaqueness of the interstellar medium. However, \citet{kap90} show
  that these emission lines originate close to the compact object,
  causing the orbital modulation of the Raman scattered
  lines. Although the central wavelength of the input EUV emission
  lines is known, they could not be identified due to the multitude of
  candidate line features in this part of the spectrum.

  The X-ray spectrum of \object{4U~1700$-$37} can be modelled as a power law
  with a high-energy cutoff \citep{hab89,rey99}, modified by an
  absorption column which depends on the orbital phase of the
  system. A sharp increase in absorption at late orbital phases
  \citep{mas76,hab89} is explained by the presence of a region of
  enhanced density trailing the compact object in its orbit, such as a
  photo-ionisation wake \citep{blo90}. The presence of such a wake in
  \object{4U~1700$-$37} has been confirmed with optical spectroscopy
  \citep{kap94}. A similar increase in absorption has been reported
  for other HMXBs, like \object{Vela~X$-$1} and \object{Cen~X$-$3} \citep{hab90,nag92}.
  \begin{figure}[!ht]
  \centering
  \includegraphics[width=8.5cm]{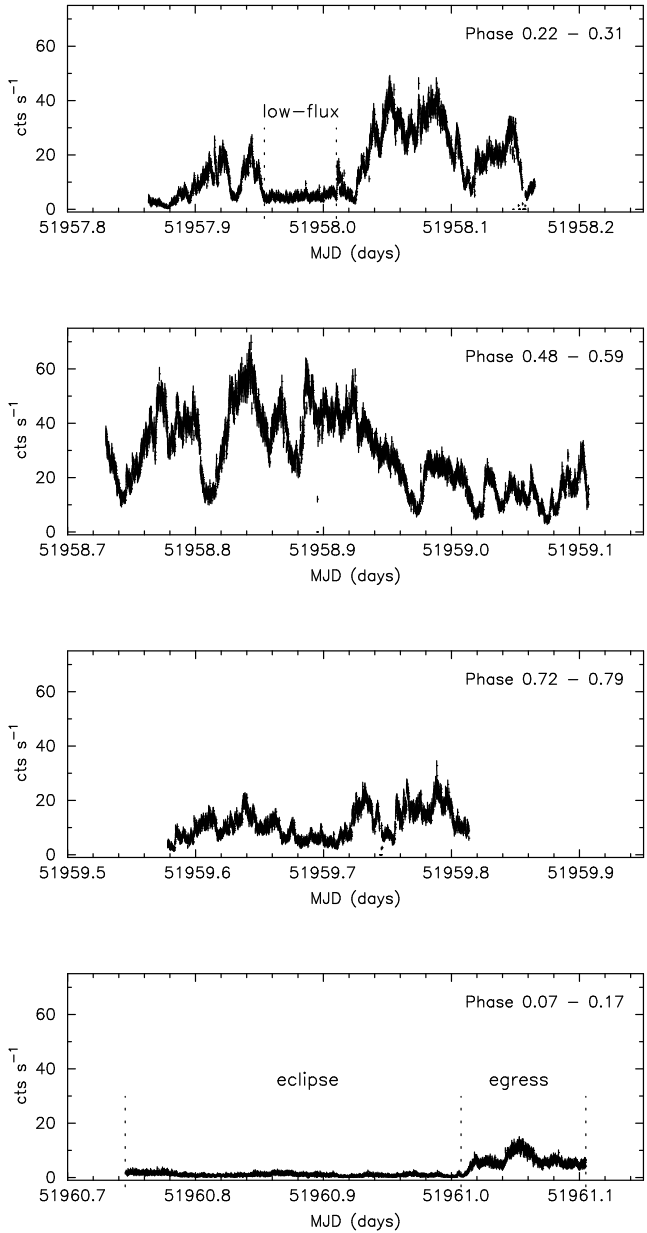}
  \caption{0.3--11.5 keV lightcurve of 4U~1700$-$37 obtained
    with the EPIC/MOS-2 camera at the end of February 2001 in four
    orbital phase intervals. The data have been extracted from a
    circular region centered on the source and binned in time-bins of
    10 seconds. The data have not been corrected for pile-up, which
    mainly distorts the shape of the spectrum but may also affect the
    value of the high count rates, e.g. during flares. We selected
    three intervals in the lightcurve (indicated by the vertical dashed
    lines), in which the spectrum does not suffer from pile-up,
    i.e. the eclipse, the eclipse egress and a low-flux interval,
    centered on orbital phase $\phi \sim 0.25$.}
  \label{lightcurve}
  \end{figure}

  A soft excess was found in {\it Ginga} observations \citep{hab92} and
  modelled as a bremsstrahlung component with a temperature of kT
  $\sim 0.5$~keV. In a second set of {\it Ginga} observations
  \citet{hab94} also found a difference in temperature before and
  after eclipse, which they tentatively explain with a bow shock in
  front of the compact object, heating the stellar wind to a higher
  temperature.

  \citet{bor03} obtained {\it Chandra} observations of \object{4U~1700$-$37} in
  August 2000 in the orbital phase interval $0.65 \le \phi \le
  0.8$. They find emission lines in the 4--13~\AA\ range, which are
  identified with fluorescence emission lines from near-neutral species
  and recombination lines from He- and H-like species.
  The X-ray source was observed during an intermittent flaring state,
  in which the strengths of the lines vary. The triplet structure of
  \ion{Si}{xiii} and \ion{Mg}{xi} is resolved but weak, but could be
  used to tentatively identify the plasma as a low-density hybrid of a
  photo-ionised and a collisionally ionised plasma \citep{bor03}.

  We monitored \object{4U~1700$-$37} with XMM-{\it Newton} around the four
  quadratures in the orbit, including part of the eclipse and the
  eclipse egress. The goal of our observing campaign is to study the
  X-ray spectrum as a function of orbital phase and to derive
  information on the physical origin of the continuum and line
  emission.
  In Sect.~\ref{observations} we describe the observations and the
  techniques that we used to reduce the dataset; in
  Sect.~\ref{spectra} we present the spectral analysis, and in
  Sect.~\ref{discussion} we discuss our results.
  \begin{table}
  \caption[]{Log of XMM-{\it Newton} observations of 4U~1700$-$37 in
    the period 17--20 February 2001. To calculate the orbital phase we
    used the ephemeris of \citet{rub96}: orbital period P$_{\rm
    orb}$ = 3.411581(27)~d and mid-eclipse T$_{0}$ = JD 2448900.873(2).}
  \begin{tabular}{cccc}
  \hline \hline
  T$_{\rm start}$ & T$_{\rm end}$ & $\Delta t$ & $\Delta \phi$ \\
  (MJD)     & (MJD)     & (ks)  & \\
  \hline
  51957.863 & 51958.166 & 25.79 & 0.22 - 0.31 \\
  51958.729 & 51959.106 & 32.87 & 0.48 - 0.59 \\
  51959.577 & 51959.813 & 20.90 & 0.72 - 0.79 \\
  51960.745 & 51961.104 & 31.36 & 0.07 - 0.17 \\
  \hline
  \end{tabular}
  \label{log}
  \end{table}

%
%

\section{Observations}
\label{observations}
  We have obtained four 20--30 ks observations of \object{4U~1700$-$37} with
  XMM-{\it Newton} \citep{jan01} in the period 2001, February
  17--20. The log of observations is listed in
  Table~\ref{log}. Fig.~\ref{lightcurve} shows the lightcurve of the
  four intervals obtained with EPIC/MOS-2 \citep{tur01}. To determine
  the orbital phase we used the ephemeris of \citet{rub96}. The X-ray
  flux of \object{4U1700$-$37} is strongly variable (factor $\sim
  10$). Numerous flares are observed with a typical duration of the
  order of half an hour.

  Due to the brightness of the optical companion (M$_{\rm V} = 6.5$),
  we were forced to switch off the Optical Monitor. The EPIC/MOS-1 and
  EPIC/PN detectors \citep{str01} were used in timing and burst mode,
  respectively, to cope with the high count rate of the X-ray
  source. EPIC/MOS-2 was used in small-window mode. For the RGS
  gratings \citep{her01} default settings were used.

  We reduced the data using the Science Analysis System (SAS) version
  5.3.3. Since this version does not yet contain a developed task to
  reduce EPIC/MOS-1 timing and EPIC/PN burst mode, in this paper we
  will concentrate on the EPIC/MOS-2 and RGS datasets. The results from 
  the EPIC/PN and EPIC/MOS-1 will be reported in a forthcoming paper.

  Since the pipeline products were created when the reduction software
  was still in an early stage, we created new event lists for the
  EPIC/MOS-2 data. These event lists were filtered following the
  data analysis threads provided by the XMM-{\it Newton} team. Because
  the instrument was set in small-window mode (to reduce the read-out
  time) our strong source produced numerous events over almost the
  whole frame, which prevented us from measuring the
  background. Therefore, we extracted a background spectrum using
  special event lists provided by the XMM-{\it Newton}
  team. \footnote{http://xmm.vilspa.esa.es/external/xmm\_sw\_cal}
  These are long exposure observations of fields with a typical
  background and without any bright sources. Since our source is
  bright, the background contribution is negligible ($\sim 0.1 \%$
  in the energy range 0.3--11.5 keV and $\lesssim 2 \%$ in the
  energy range 0.3--2.0 keV).
  \begin{figure*}[!t]
  \centering
  \resizebox{16cm}{!}{\includegraphics[angle=-90]{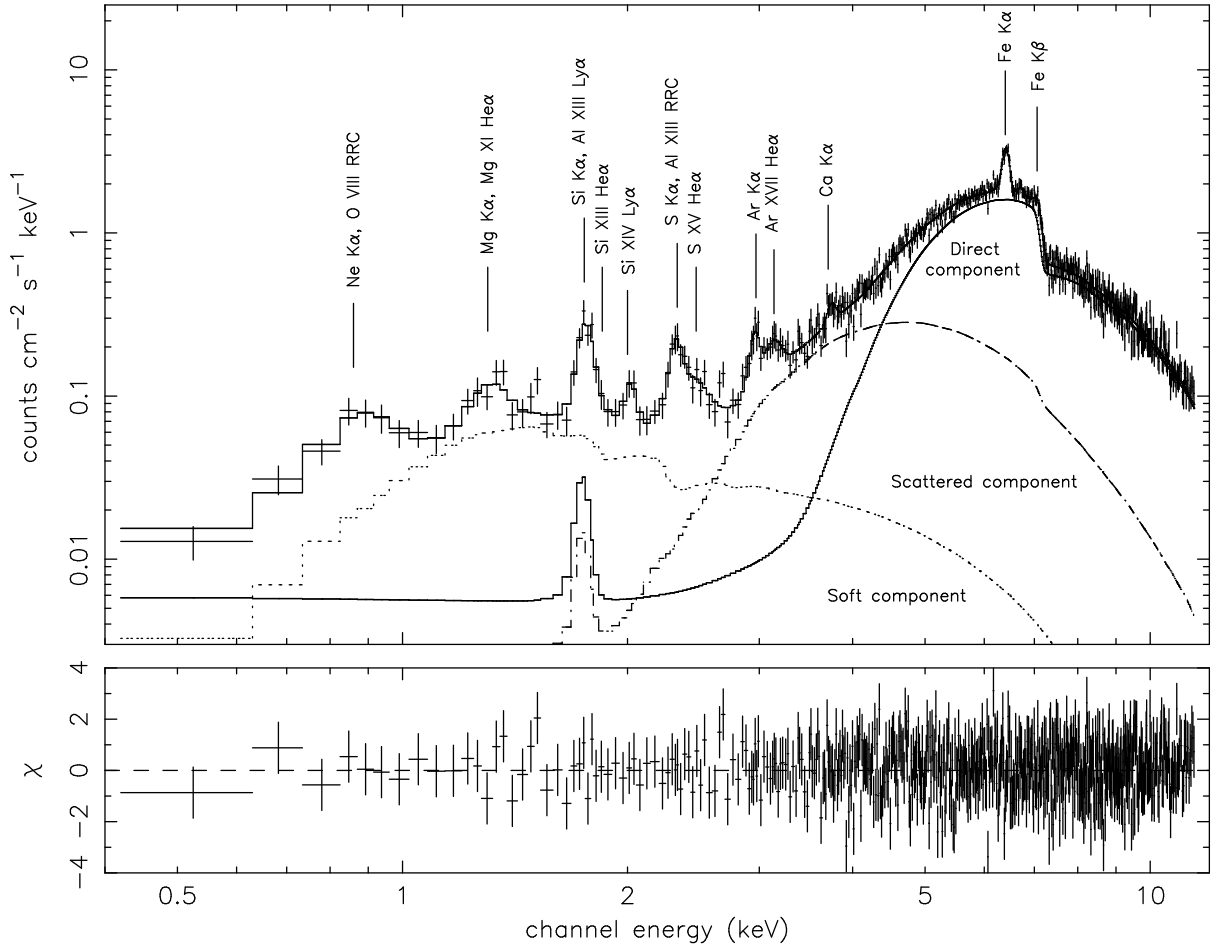}}
  \caption{XMM/EPIC observation of 4U~1700$-$37 during eclipse egress
    ($0.14 \le \phi \le 0.17$). The continuum emission is modelled as
    three power laws, representing a direct (solid line), a scattered
    (dashed-dotted line) and a soft component (dotted line). The
    spectrum clearly shows several fluorescence emission lines and
    some discrete recombination lines of He- and H-like species (see
    Table~\ref{linetable} for their identification). Note that the
    data in this plot are folded with the response matrix of the
    instrument. Therefore, some of the features (e.g. at 1.8~keV) in
    the plot are of instrumental origin. The lower panel shows the
    residuals of the model to the spectrum.}
  \label{egress}
  \end{figure*}

  Due to the high count rate a large fraction of the dataset is 
  affected by pile-up. The basic concept of pile-up is that a
  saturation effect takes place in which two or more
  separate photons fall onto the same pixel within a read-out cycle
  and are counted as one with a total energy equal to the
  sum of their energies (for a detailed description see
  e.g. \citealt{bal99}). This effect distorts the source spectrum and
  the measured flux. The SAS software provides the tool EPATPLOT to
  analyse the distribution of patterns produced on the CCDs by the
  incoming photons. Two or more photons counted as one have a lower
  probability to produce a single-pixel pattern than one photon with
  the same energy. This affects the observed pattern distribution,
  which can be compared with the expected distribution to assess the
  pile-up. However, with this method it is not clear to what extent
  the spectral parameters (derived from fits to the extracted
  spectrum) are affected by pile-up.

  In order to assess the magnitude of the distortion and to eliminate
  the part of the point spread function (PSF) affected by pile-up, we
  created multiple event lists, extracted from annuli centered on the
  source, with increasing inner radii (with steps of $2.5\arcsec$) up
  to a maximum of $25\arcsec$. We modelled the spectra with
  the same basic model of two absorbed power laws (see
  Sect.~\ref{sect_continuum}). The best-fitted spectral parameters were
  then examined as a function of inner radius of the extraction
  region. One expects the parameters to stabilise beyond the radius at
  which pile-up no longer affects the remaining part of the
  PSF. However, we find that the parameters continue changing,
  sometimes even up to the maximum inner radius, while EPATPLOT
  already gives satisfactory results. We conclude that the model of
  the energy dependence of the PSF is not known accurately enough for
  our purpose.

  In the current study we concentrate on data with a sufficiently low
  countrate. We select three intervals in the lightcurve (see
  Fig.~\ref{lightcurve}): eclipse, eclipse egress and a low-flux part
  centered on orbital phase $\phi \sim 0.25$. These intervals of low
  flux are not uncommon in \object{4U~1700$-$37} \citep[e.g.][]{bor03} and are
  also observed in other HMXBs, e.g. \object{4U~1907$+$097} studied by
  \citet{zan97}, who refer to this behaviour as ``dipping activity''.

  We also analysed the RGS data extracted in the same
  timeframes. However, the source is highly absorbed, resulting in a
  very low RGS-countrate. After rebinning the data to 20-25 counts per
  bin the RGS spectra were used to verify the best-fit models for the
  EPIC/MOS-2 data.

%
%

\section{Spectral Analysis}
\label{spectra}
  The EPIC/MOS-2 camera covers the energy range 0.3--11.5~keV. In
  previous studies based on a wider energy range (EXOSAT, 3--20~keV,
  \citealt{hab89}; {\it Beppo}SAX, 0.5--200~keV, \citealt{rey99}) the
  continuum of \object{4U~1700$-$37} is modelled as an absorbed power law with
  photon index $\alpha$, modified above a break energy $E_{c}$ by a
  high-energy cutoff of the form $\exp{ \left[ (E_{c} - E)/E_{f}
  \right]}$.
  \begin{figure*}[!ht]
  \centering
  \resizebox{16cm}{!}{\includegraphics[angle=-90]{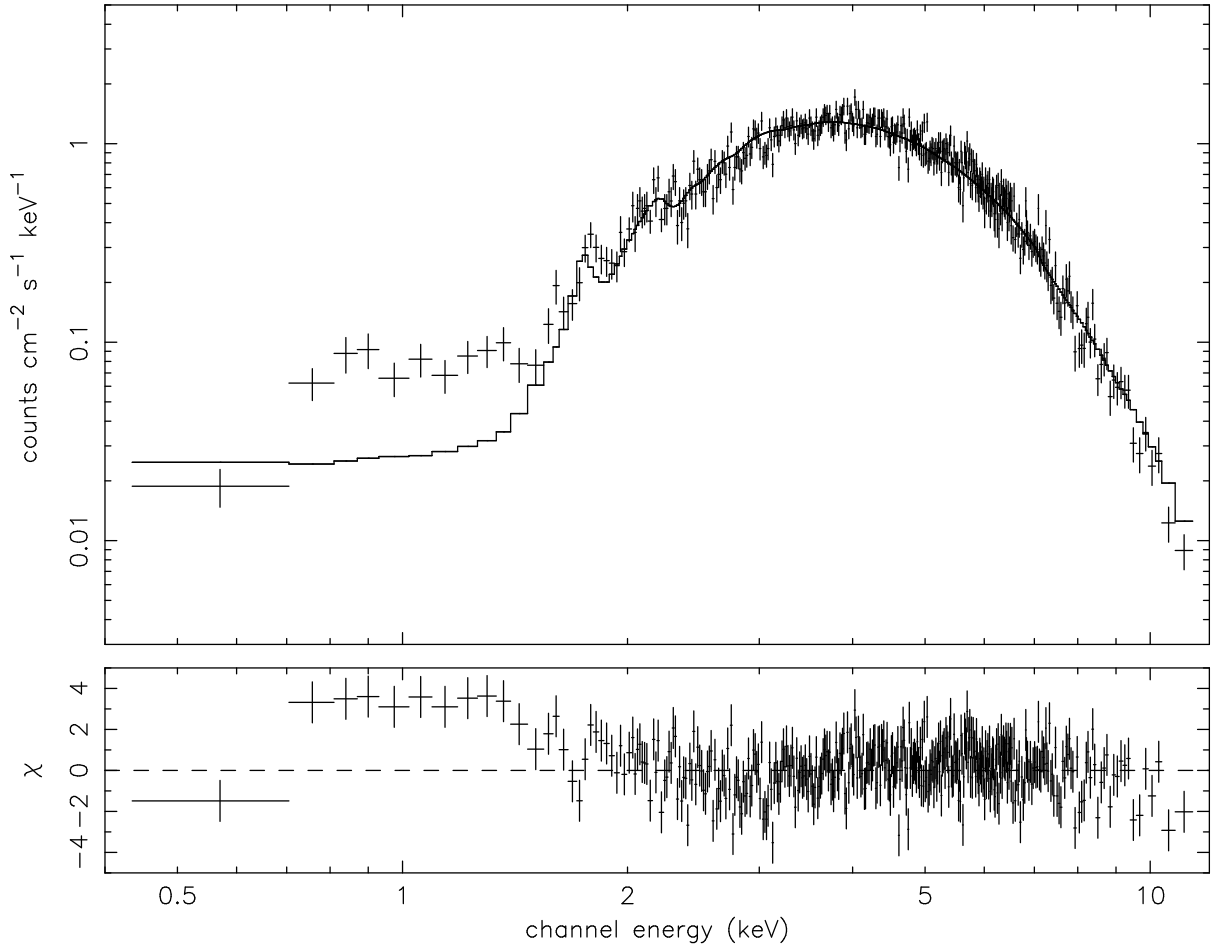}}
  \caption{XMM/EPIC observation of the 4U~1700$-$37 during the
    low-flux interval. A soft excess below $\sim 2$~keV is clearly
    present. The solid line represents a model that consists of one absorbed
    power-law component. The lower panel shows the residuals of the model to the
    spectrum.}
  \label{low-flux_single}
  \end{figure*}

  \citet{sak99} obtain a good fit of the ASCA eclipse spectrum of
  \object{Vela~X$-$1} using two absorbed power laws with identical photon
  index, but different column density $N_{H}$ and normalisation to
  model the continuum. One power law describes the direct component,
  that originates from close to the neutron star. A part of this
  component is scattered by the extended stellar wind of the optical
  companion and is described by the second powerlaw. \citet{bor03}
  apply the same method to fit the {\it Chandra} spectrum of
  \object{4U~1700$-$37}. We follow the same strategy modelling the XMM-{\it
  Newton} spectra. We apply three models to fit the continuum at
  eclipse, egress and low flux. The emission lines are subsequently
  fit by gaussians. The spectra have been analysed using XSPEC version
  11.2.0, after rebinning the spectra to 25 counts per bin.

%
%

  \subsection{Continuum Emission}
  \label{sect_continuum}
  We first describe the eclipse egress spectrum, where a clear
  distinction can be made between the different continuum components
  (see Fig.~\ref{egress}). We discuss the high-energy and the
  low-energy part of the spectra separately. Note that the mentioned
  parameters correspond to the models fitted to the full 0.3--11.5 keV
  energy range. Table~\ref{continua} lists the models we used to fit
  these three spectra of which the parameters are constrained:
  model~A, which consists of three power laws, model~B with two (or
  three) power laws modified by a high-energy cutoff, and model~C,
  similar to model~A, but with a blackbody component replacing the
  power law describing the soft component.

%
%

  \subsubsection{Direct and scattered component}
  We model the high energy part of the eclipse egress spectrum as two
  power laws, each of them modified by different absorbing columns
  \citep{mor83} depending on the orbital phase of the system (for the
  egress spectrum these are N$_{\rm H} \sim 8.3 \times 10^{23}$
  cm$^{-2}$ and N$_{\rm H} \sim 1.8 \times 10^{23}$ cm$^{-2}$; see
  model~A in Table~\ref{continua}). Both components have the same
  power law index  $\alpha \sim 1.4$. The hard, direct component is
  assumed to originate from the near surroundings of the compact
  object and the scattered component is produced by Thomson scattering
  by electrons in the extended stellar wind of the optical
  companion. Since Thomson scattering is energy independent, the
  photon index of the scattered component is expected to be the same, and
  is therefore fixed to the index of the direct component. The
  best-fit values of the model parameters are given in
  Table~\ref{continua}. Both components are also detected during the
  part of the eclipse covered by our observations ($0.07 \le \phi \le
  0.14$), although the flux of the direct component is reduced by a
  factor $\sim 45$ compared to egress.
  \begin{sidewaystable*}[!hp]
  \caption[]{Parameters used to describe the continuum models of the
    eclipse egress, eclipse and low-flux spectra. All errors and
    upper limits are a 1 $\sigma$ confidence level. Three different models are
    considered. Model~A consists of three power-law components with identical
    photon index, but different normalisation, which are modified by
    different absorption columns. Model~B contains two power-law
    components, modified by a high-energy cutoff. Model~C
    includes a blackbody component replacing one of the three
    power-law components of model~A.}
  \label{continua}
  \begin{flushleft}
  \begin{tabular}{p{7.5cm}|p{1.4cm}*{2}{p{1.4cm}}|p{1.4cm}*{2}{p{1.4cm}}|p{1.4cm}*{2}{p{1.4cm}}}
  \hline \hline
  & \multicolumn{3}{|c|}{\rule[-2mm]{0mm}{6mm}egress} 
  & \multicolumn{3}{|c|}{eclipse} 
  & \multicolumn{3}{|c}{low-flux} \\ \hline
  & \multicolumn{1}{|l}{Model A} 
  & \multicolumn{1}{l}{Model B}
  & \multicolumn{1}{l|}{Model C} 
  & \multicolumn{1}{|l}{Model A} 
  & \multicolumn{1}{l}{Model B}
  & \multicolumn{1}{l|}{Model C} 
  & \multicolumn{1}{|l}{Model A}
  & \multicolumn{1}{l}{Model B}
  & \multicolumn{1}{l}{Model C} \\
  \hline 
  \vspace{-0.1cm}
  {\it direct component}                                                          &                  &                  &                             &                  &                  &                             &                             &                             &                  \\[0.05cm]
  N$_{\rm H}$ ($\times 10^{22}$ cm$^{-2}$)                                        & $83 \pm 2      $ & $85 \pm 3      $ & $83 \pm 2                 $ & $59 \pm 8      $ & $89 \pm 12     $ & $61 \pm 9                 $ & $20 \pm 3                 $ & $6.8 \pm 0.4              $ & $17 \pm 2      $ \\[0.05cm]
  Normalisation ($\times 10^{-2}$ photons s$^{-1}$ cm$^{-2}$ keV$^{-1}$ at 1 keV) & $44 \pm 6      $ & $12 \pm 2      $ & $44 \pm 6                 $ & $1.1 \pm 0.4   $ & $0.47 \pm 0.19 $ & $1.1  _{- 0.3  }^{+ 0.6  }$ & $8.7 \pm 2.4              $ & $3.8 \pm 0.6              $ & $9.3 \pm 2.0   $ \\[0.05cm]
  Unabsorbed flux ($\times 10^{-10}$ erg s$^{-1}$ cm$^{-2}$)                      & $49 \pm 3      $ & $34 \pm 5      $ & $49 \pm 3                 $ & $1.1 \pm 0.3   $ & $1.0 \pm 0.4   $ & $1.2 \pm 0.4              $ & $5.3  _{- 0.7  }^{+ 1.5  }$ & $5.3 \pm 1.0              $ & $6.1 \pm 1.0   $ \\[0.25cm]
  \hline
  \vspace{-0.1cm}
  {\it scattered component}                                                       &                  &                  &                             &                  &                  &                             &                             &                             &                  \\[0.05cm]
  N$_{\rm H}$ ($\times 10^{22}$ cm$^{-2}$)                                        & $17.5 \pm 1.3  $ & $18.7 \pm 1.6  $ & $17.6 \pm 2.2             $ & $11.6 \pm 0.7  $ & $11.2 \pm 0.7  $ & $12.3 \pm 1.0             $ & $7.2 \pm 0.3              $ & $...                      $ & $6.4 \pm 0.4   $ \\[0.05cm]
  Normalisation ($\times 10^{-2}$ photons s$^{-1}$ cm$^{-2}$ keV$^{-1}$ at 1 keV) & $1.6 \pm 0.3   $ & $0.67 \pm 0.13 $ & $1.6 \pm 0.4              $ & $0.56 \pm 0.12 $ & $0.24 \pm 0.05 $ & $0.61 \pm 0.15            $ & $8.2 \pm 1.4              $ & $...                      $ & $6.3 \pm 1.4   $ \\[0.05cm]
  Unabsorbed flux ($\times 10^{-10}$ erg s$^{-1}$ cm$^{-2}$)                      & $1.7 \pm 0.2   $ & $1.9 \pm 0.4   $ & $1.8 \pm 0.3              $ & $0.60 \pm 0.06 $ & $0.54 \pm 0.12 $ & $0.66 \pm 0.07            $ & $5.3 \pm 0.7              $ & $...                      $ & $4.1 \pm 0.7   $ \\[0.25cm]
  \hline
  \vspace{-0.1cm}
  {\it soft component}                                                            &                  &                  &                             &                  &                  &                             &                             &                             &                  \\[0.05cm]
  N$_{\rm H}$ ($\times 10^{22}$ cm$^{-2}$)                                        & $0.56 \pm 0.09 $ & $0.36 \pm 0.08 $ & $0.30 \pm 0.17            $ & $0.24 \pm 0.04 $ & $0.13 \pm 0.04 $ & $<0.05                    $ & $0.43 \pm 0.08            $ & $0.32 \pm 0.07            $ & $1.0 \pm 0.4   $ \\[0.05cm]
  Normalisation ($\times 10^{-4}$ photons s$^{-1}$ cm$^{-2}$ keV$^{-1}$ at 1 keV) & $4.3 \pm 0.5   $ & $2.6 \pm 0.3   $ & $...                      $ & $2.3 \pm 0.2   $ & $1.7 \pm 0.2   $ & $...                      $ & $4.7 \pm 0.9              $ & $3.4 \pm 0.7              $ & $...           $ \\[0.05cm]
  Blackbody temperature (keV)                                                     & $...           $ & $...           $ & $0.74 _{- 0.19 }^{+ 0.34 }$ & $...           $ & $...           $ & $0.77 _{- 0.03 }^{+ 0.11 }$ & $...                      $ & $...                      $ & $0.13 \pm 0.04 $ \\[0.05cm]
  Normalisation ($\times 10^{32}$ erg s$^{-1}$ kpc$^{-2}$)                        & $...           $ & $...           $ & $2.1  _{- 0.5  }^{+ 1.7  }$ & $...           $ & $...           $ & $1.3 \pm 0.4              $ & $...                      $ & $...                      $ & $<270          $ \\[0.05cm]
  Unabsorbed flux ($\times 10^{-12}$ erg s$^{-1}$ cm$^{-2}$)                      & $4.7 \pm 0.5   $ & $7.5 \pm 0.8   $ & $1.8  _{- 0.4  }^{+ 1.4  }$ & $2.5 \pm 0.3   $ & $3.7 \pm 0.4   $ & $1.1 \pm 0.3              $ & $3.0 \pm 0.5              $ & $4.7 \pm 1.1              $ & $<230          $ \\[0.25cm]
  \hline
  &&&&&&& \\
  Cutoff energy (keV)                                                             & $...           $ & $5.4 $ fixed     & $...                      $ & $...           $ & $5.4 $ fixed     & $...                      $ & $...                      $ & $5.4 \pm 0.4              $ & $...           $ \\[0.05cm]
  Folding energy (keV)                                                            & $...           $ & $10.6 $ fixed    & $...                      $ & $...           $ & $10.6 $ fixed    & $...                      $ & $...                      $ & $10.6 _{- 1.6  }^{+ 2.3  }$ & $...           $ \\[0.25cm]
  Photon index                                                                    & $1.35 \pm 0.05 $ & $0.61 \pm 0.05 $ & $1.35 \pm 0.05            $ & $1.36 \pm 0.12 $ & $0.77 \pm 0.12 $ & $1.37 \pm 0.13            $ & $1.87 _{- 0.04 }^{+ 0.09 }$ & $1.08 \pm 0.09            $ & $1.83 \pm 0.08 $ \\[0.25cm]
  $\chi^{2} / \rm{d.o.f}$                                                         & $558/522       $ & $578/522       $ & $559/521                  $ & $444/408       $ & $445/408       $ & $443/407                  $ & $485/390                  $ & $487/390                  $ & $475/389       $ \\[0.05cm] 
  \hline
  \end{tabular}
  \end{flushleft}
  \end{sidewaystable*}

  In the low-flux spectrum it is difficult to separate the two
  components. The model produces comparable strengths and absorption
  coefficients (Model A in Table~\ref{continua}) with a photon index
  that is slightly higher ($\alpha \sim 1.9$) than in eclipse and
  egress.
  Modelling the spectrum with only one power law modified by a
  high-energy cutoff gives an equally good fit (model~B in
  Table~\ref{continua}). The resulting photon index is lower ($\alpha
  \sim 1.1$) than in model~A. The cutoff energy in this model is $\sim
  5.4$~keV (see Table~\ref{continua}), which is in agreement with
  previous observations (e.g. \citealt{hab89}, who find $E_{c} = 6.6
  \pm 0.7$~keV in EXOSAT observations and \citealt{rey99} who find $E_{c}
  = 5.9 \pm 0.2$~keV in {\it Beppo}SAX observations). However, the
  folding energy we find ($E_{f} \sim 11$~keV) is low; \citet{hab89}
  find $E_{f} = 21 \pm 4$~keV and \citet{rey99} measure $E_{f} = 23.9
  \pm 0.9$~keV. 
  \begin{figure*}[!ht]
  \centering
  \resizebox{16cm}{!}{\includegraphics[angle=-90]{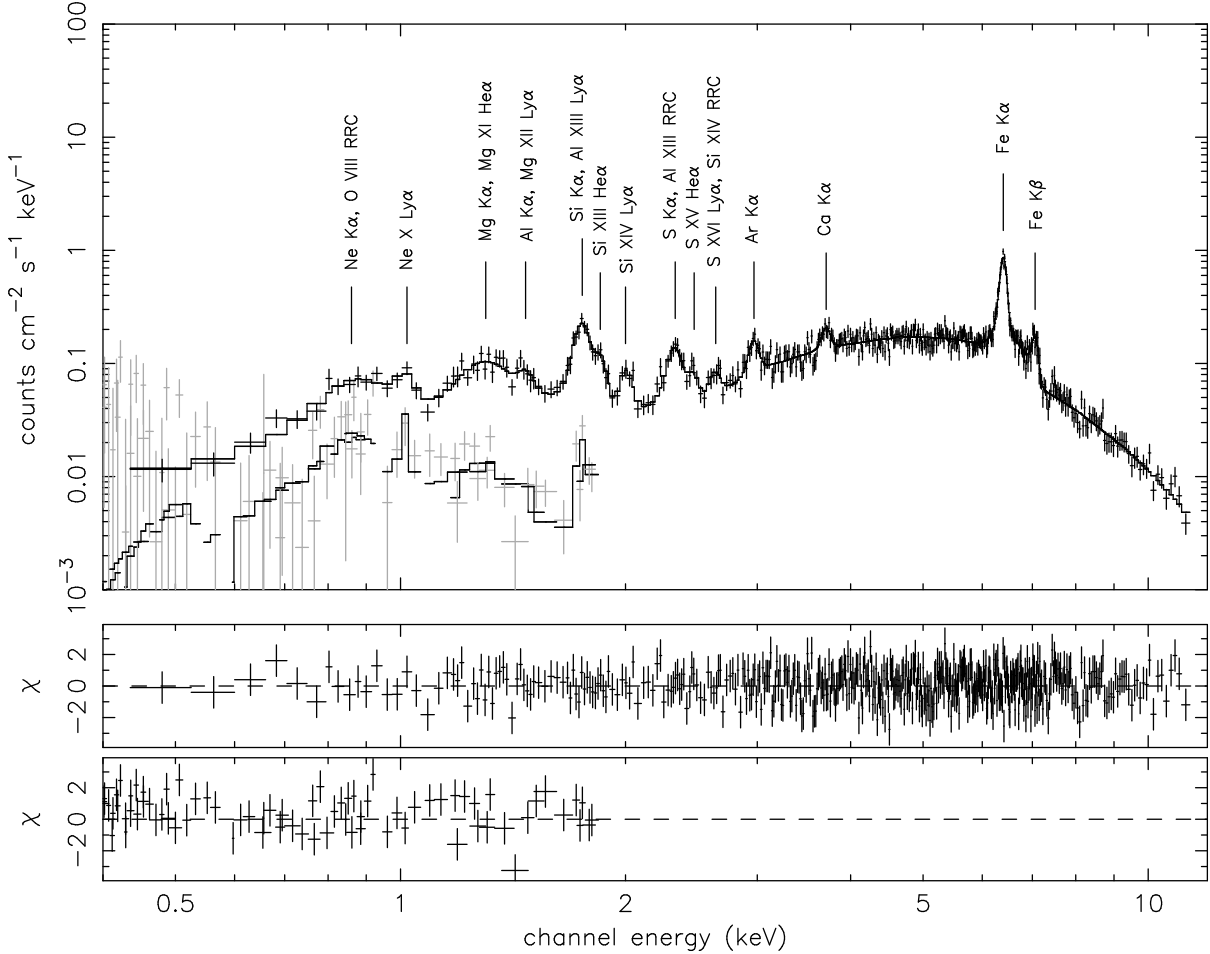}}
  \caption{XMM/EPIC observation of 4U~1700$-$37 during eclipse. The
    spectrum clearly shows several fluorescence emission lines and
    some discrete recombination lines of He- and H-like species with
    their identifications. The continuum model used for this
    spectrum is model~A (see Table~\ref{continua}); the RGS spectrum is
    also included. The EPIC/MOS-2 residuals are shown in the middle
    panel, and the RGS residuals in the lower panel.}
  \label{eclipse}
  \end{figure*}

  The problem with applying this model to our dataset is
  that the folding energy is outside the energy range of EPIC/MOS-2
  (0.3--11.5~keV) and its value is easily influenced by the value of
  the photon index. If we apply the model to the eclipse and egress
  spectrum, a high-energy cutoff yields an unacceptable fit, and the
  fit parameters can not be constrained. If we fix E$_{c}$ and E$_{f}$
  to the values obtained from the low-flux spectrum, the fit
  parameters can be constrained (see Table~\ref{continua}). This is
  also the case if we fix their values to the values of the {\it
  Beppo}SAX observations \citep{rey99}. The fits with model~A and with
  this ``fixed'' version of model~B are of comparable quality for the
  eclipse spectrum, but for egress fits with model~B are worse than
  fits with model~A ($\chi^{2}/{\rm d.o.f.} = 578/522$ for model~B
  compared to $558/522$ for model~A; d.o.f. = degrees of
  freedom). Note that model~B also indicates a higher photon index in
  the low-flux interval compared to eclipse and egress.

%
%

  \subsubsection{Soft Excess}
  After having obtained the best fit for the models discussed above, an
  excess remains in all cases in the low energy part of the spectrum
  (see Fig.~\ref{low-flux_single}). A soft excess is reported for many
  X-ray spectra of HMXBs, but the nature of this component is not
  clear. To investigate the physical origin of this component
  we fit the soft component with four different models, which are
  discussed below: bremsstrahlung, blackbody, electron scattering and
  dust scattering. In all cases it is found that the model component
  needs to be modified by a moderate absorption column, close to the
  interstellar value N$_{\rm H} \sim 0.3 \times 10^{22}$ cm$^{-2}$ for
  this source, based on the observed reddening $\rm E(B-V) = 0.59$
  \citep{hut73} and N$_{\rm H} \rm /E(B-V) = 0.58 \times 10^{22}$
  cm$^{-2}$ mag$^{-1}$ \citep{kil92}.

  Adding a bremsstrahlung component, as proposed by \citet{hab92} and
  \citet{hab94}, improves the fit to all three spectra significantly
  ($\chi^{2}/{\rm d.o.f.} = 476/389$ for the low-flux
  spectrum). However, the temperature of the bremsstrahlung component
  can not be constrained in any of the spectra. Actually, this
  bremsstrahlung component tends to mimic the shape of a power
  law. This suggests that bremsstrahlung is not the physical origin of
  the soft excess.

  Fitting this excess with a blackbody component gives a similar improvement
  of the fit ($\chi^{2}/{\rm d.o.f.} = 475/389$ for the low-flux
  spectrum). For this model the temperature $\rm kT \sim
  0.13$~keV. The flux is not constrained in the low-flux spectrum, but
  we can place an upper limit of $2.3 \times 10^{-10}$ erg s$^{-1}$
  cm$^{-2}$. In the eclipse and egress spectra the resulting
  temperature is slightly higher ($\rm kT \sim 0.77$~keV and $\rm kT
  \sim 0.74$~keV, respectively). The unabsorbed flux is constrained to
  $1.1 \times 10^{-12}$ erg s$^{-1}$ cm$^{-2}$ and $1.8 \times
  10^{-12}$ erg s$^{-1}$ cm$^{-2}$, respectively (see model~C in
  Table~\ref{continua}).

  Modelling the soft excess with yet another (third) power law with a
  photon index fixed to that of the direct and scattered component
  also gives a significant improvement ($\chi^{2}/{\rm d.o.f.} = 485/390$) and all
  the parameters are constrained (see Table~\ref{continua}). This
  component could, like the scattered component of model~A, represent a
  Thomson scattered version of the hard, direct component, with a very
  low absorption column (N$_{\rm H} \sim 0.43 \times 10^{22}$
  cm$^{-2}$).
  \begin{table*}[t]
  \caption[]{List of the emission lines that have been detected in the
  eclipse, the eclipse-egress and the low flux spectra. All errors and
  upper limits are 1 $\sigma$. If a line is detected in one spectrum
  but not in another, an upper limit on the flux is given.}
  \label{linetable}
  \begin{flushleft}
  \begin{tabular}{p{4.5cm}*{3}{p{1.5cm}}p{0.5cm}*{3}{p{1.5cm}}}
  \hline \hline
  & eclipse & egress & low-flux & & eclipse & egress & low-flux\\
  \hline \\
  Energy (keV)                                & $0.86 \pm 0.02$        & $0.87 \pm 0.02$        & $0.86 \pm 0.02$ & & $2.46 \pm 0.01$        & $2.42_{-0.02}^{+0.04}$ & $2.46$ fixed                 \\[0.05cm]
  Width (eV)                                  & $71 \pm 19$            & $84 \pm 32$            & $60 \pm 27$     & & $<18$                  & $115 \pm 29$           & $20$ fixed                   \\[0.05cm]
  Flux ($10^{-5}$ photons cm$^{-2}$ s$^{-1}$) & $4.0 \pm 1.0$          & $5.5 \pm 2.0$          & $5.3 \pm 1.8 $  & & $1.6 \pm 0.4$          & $9.8 \pm 2.9$          & $<6.1$                       \\[0.05cm]
  Identification                              & $\ion{Ne~K\alpha}{},~\ion{O}{viii}~{\rm RRC}$ & &                 & & $\ion{S}{xv}~{\rm He\alpha}$ &                  &                              \\[0.25cm]
  Energy (keV)                                & $1.01 \pm 0.01$        & $1.01$ fixed           & $1.01$ fixed    & & $2.63 \pm 0.01$        & $2.63$ fixed           & $2.63$ fixed                 \\[0.05cm]
  Width (eV)                                  & $<13$                  & $13$ fixed             & $13$ fixed      & & $<27$                  & $27$ fixed             & $27$ fixed                   \\[0.05cm]
  Flux ($10^{-5}$ photons cm$^{-2}$ s$^{-1}$) & $1.0 \pm 0.3$          & $<0.52$                & $<1.7$          & & $1.1 \pm 0.3$          & $<1.8$                 & $<2.0$                       \\[0.05cm]
  Identification                              & $\ion{Ne}{x}~{\rm Ly\alpha}$ &                  &                 & & $\ion{S}{xvi}~{\rm Ly\alpha},~\ion{Si}{xiv}~{\rm RRC}$ & &                     \\[0.25cm]
  Energy (keV)                                & $1.30 \pm 0.01$        & $1.32 \pm 0.03$        & $1.31$ fixed    & & $2.97 \pm 0.01$        & $2.96 \pm 0.01$        & $2.97$ fixed                 \\[0.05cm]
  Width (eV)                                  & $77 \pm 17$            & $<87$                  & $87$ fixed      & & $<29$                  & $<38$                  & $29$ fixed                   \\[0.05cm]
  Flux ($10^{-5}$ photons cm$^{-2}$ s$^{-1}$) & $3.1 \pm 0.7$          & $2.2 \pm 0.7$          & $<3.3$          & & $2.8 \pm 0.4$          & $4.7 \pm 0.9$          & $<4.6$                       \\[0.05cm]
  Identification                              & $\ion{Mg~K\alpha}{},~\ion{Mg}{xi}~{\rm He\alpha}$ & &             & & $\ion{Ar~K\alpha}{}$   &                        &                              \\[0.25cm]
  Energy (keV)                                & $1.47 \pm 0.02$        & $1.47$ fixed           & $1.47$ fixed    & & $3.14$ fixed           & $3.14 \pm 0.02$        & $3.14$ fixed                 \\[0.05cm]
  Width (eV)                                  & $<31$                  & $31$ fixed             & $31$ fixed      & & $56$ fixed             & $56 \pm 32$            & $56$ fixed                   \\[0.05cm]
  Flux ($10^{-5}$ photons cm$^{-2}$ s$^{-1}$) & $0.6 \pm 0.2$          & $<0.78$                & $<0.15$         & & $<0.69$                & $4.3 \pm 1.4$          & $<0.81$                      \\[0.05cm]
  Identification                              & $\ion{Al~K\alpha}{},~\ion{Mg}{xii}~{\rm Ly\alpha}$ & &            & & $\ion{Ar}{xvii}~{\rm He\alpha}$ &               &                              \\[0.25cm]
  Energy (keV)                                & $1.75 \pm 0.01$        & $1.77 \pm 0.01$        & $1.76$ fixed    & & $3.71 \pm 0.01$        & $3.75 \pm 0.02$        & $3.71$ fixed                 \\[0.05cm]
  Width (eV)                                  & $<26$                  & $<26$                  & $26$ fixed      & & $<28$                  & $<46$                  & $30$ fixed                   \\[0.05cm]
  Flux ($10^{-5}$ photons cm$^{-2}$ s$^{-1}$) & $5.0 \pm 0.6$          & $4.8 \pm 0.9$          & $<0.35$         & & $2.7 \pm 0.5$          & $3.4 \pm 1.1$          & $<6.3$                       \\[0.05cm]
  Identification                              & $\ion{Si~K\alpha}{},~\ion{Al}{xiii}~{\rm Ly\alpha}$ & &           & & $\ion{Ca~K\alpha}{}$   &                        &                              \\[0.25cm]
  Energy (keV)                                & $1.85 \pm 0.01$        & $1.86 \pm 0.06$        & $1.83 \pm 0.03$ & & $6.40 \pm 0.01$        & $6.42 \pm 0.01$        & $6.41 _{- 0.04 }^{+ 0.05 } $ \\[0.05cm]
  Width (eV)                                  & $<20$                  & $<101$                 & $<42$           & & $<15$                  & $22 \pm 10$            & $<65                       $ \\[0.05cm]
  Flux ($10^{-5}$ photons cm$^{-2}$ s$^{-1}$) & $3.4 \pm 0.8$          & $1.6_{-0.9}^{+1.6}$    & $2.7 \pm 1.5 $  & & $62.5 \pm 1.9$         & $150 \pm 7$            & $5.3  _{- 2.1  }^{+ 7.3  } $ \\[0.05cm]
  Identification                              & $\ion{Si}{xiii}~{\rm He\alpha}$ &               &                 & & $\ion{Fe~K\alpha}{i-xvii}$ &                    &                              \\[0.25cm]
  Energy (keV)                                & $2.00 \pm 0.01$        & $2.02 \pm 0.01$        & $2.01$ fixed    & & $6.53 \pm 0.04$        & $6.72 \pm 0.02$        & $6.72$ fixed                 \\[0.05cm]
  Width (eV)                                  & $<18$                  & $<31$                  & $20$ fixed      & & $206 \pm 30$           & $<42$                  & $42$ fixed                   \\[0.05cm]
  Flux ($10^{-5}$ photons cm$^{-2}$ s$^{-1}$) & $1.6 \pm 0.3$          & $1.9 \pm 0.5$          & $<4.2$          & & $19.2 \pm 2.9$         & $20 \pm 5$             & $<0.92$                      \\[0.05cm]
  Identification                              & $\ion{Si}{xiv}~{\rm Ly\alpha}$ &                &                 & & $\ion{Fe~K\alpha}{xviii-xxiv},~\ion{Fe}{xxv}~{\rm He\alpha}$ & &               \\[0.25cm]
  Energy (keV)                                & $2.33 \pm 0.01$        & $2.32 \pm 0.01$        & $2.33$ fixed    & & $7.06 \pm 0.01$        & $7.11 \pm 0.01$        & $7.06$ fixed                 \\[0.05cm]
  Width (eV)                                  & $34 \pm 11$            & $<34$                  & $34$ fixed      & & $<28$                  & $37 \pm 25$            & $28$ fixed                   \\[0.05cm]
  Flux ($10^{-5}$ photons cm$^{-2}$ s$^{-1}$) & $6.3 \pm 0.7$          & $6.1 \pm 1.4$          & $<4.9$          & & $11.3 \pm 1.1$         & $60 \pm 7$             & $<13.1$                      \\[0.05cm]
  Identification                              & $\ion{S~K\alpha}{},~\ion{Al}{xiii}~{\rm RRC}$ & &                 & & $\ion{Fe~K\beta}{i-xvii}$ &                     &                              \\[0.25cm]
  \hline
  \end{tabular}
  \end{flushleft}
  \end{table*}

  Scattering by dust grains becomes a possibility at larger distance
  from the system (where the absorption is low) as found by
  \citet{ebi96} in the ASCA observations of \object{Cen~X$-$3}. We add 2 to the
  photon index of the soft component compared to the index of the hard
  component, since dust scattering, unlike Thomson scattering, is
  energy dependent ($\propto 1/E^{2}$). This model results in an
  equally good fit as Thomson scattering for the low-flux and egress
  spectrum. Since the difference between Thomson scattering
  and dust scattering in our models is produced by a change in the
  photon index, this will only be visible as a difference in
  flux at the high energy part of the spectrum. However, this
  part of the spectrum is dominated by the flux of the direct
  component and therefore the difference between dust or Thomson
  scattering is difficult to measure. For the eclipse spectrum, where
  the direct component does not dominate the high-energy part of the
  spectrum, the dust scattering model yields a significantly worse fit
  ($\chi^{2}/{\rm d.o.f.} = 453/408$) than Thomson scattering
  ($\chi^{2}/{\rm d.o.f.} = 444/408$).
  \begin{figure*}[!ht]
  \centering
  \resizebox{16cm}{!}{\includegraphics[angle=-90]{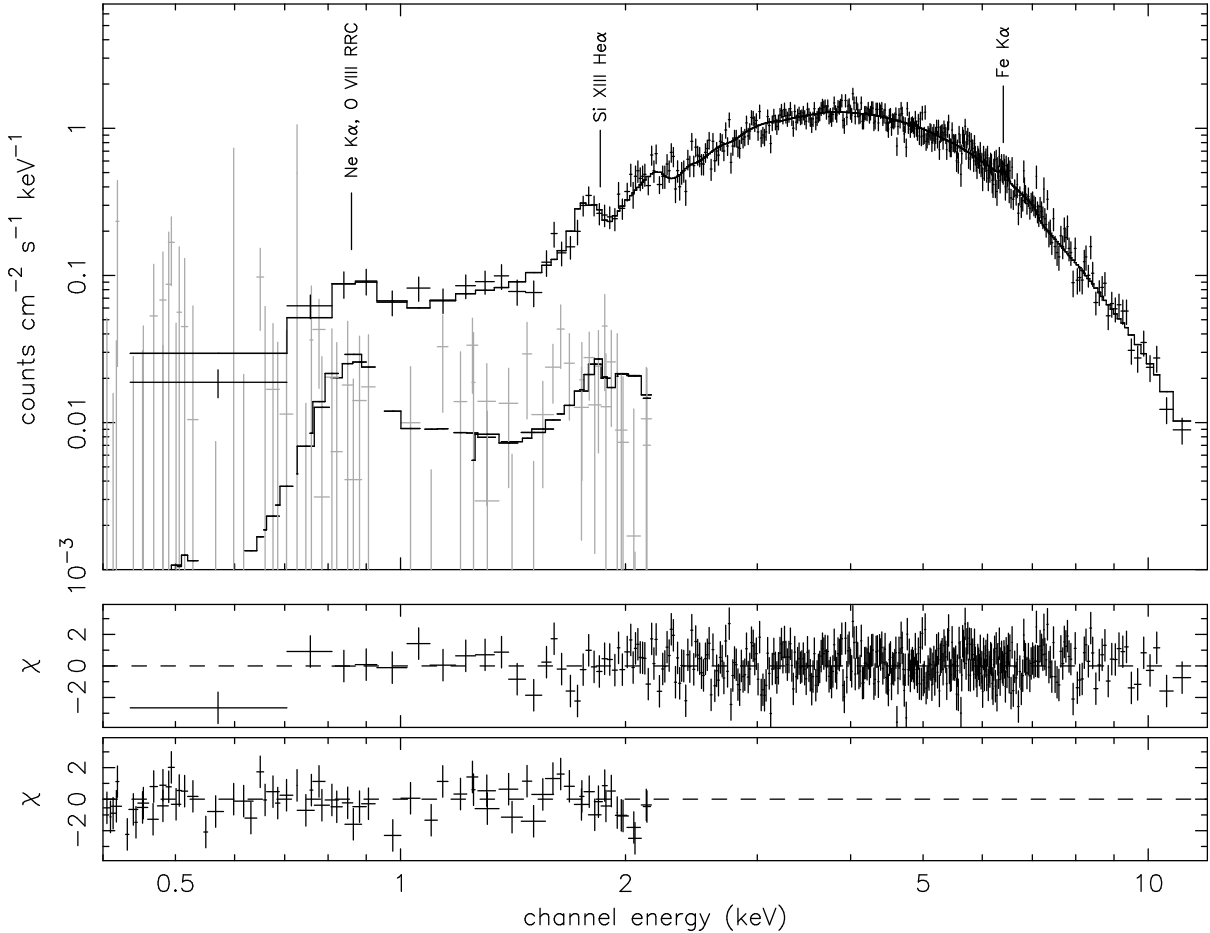}}
  \caption{XMM/EPIC observation of the 4U~1700$-$37 during low flux. The
    soft excess has been fit by a third power-law component. Only some
    of the many lines observed in eclipse and egress are
    detected. Especially the strong reduction of the iron complex is
    remarkable. The RGS spectrum is displayed in the left bottom
    corner of the upper panel. The EPIC/MOS-2 residuals are shown in
    the middle panel, and the RGS residuals in the lower panel.}
  \label{low-flux_lines}
  \end{figure*}

  Following the approach of \citet{bor03} we also try to model the
  soft excess as a blend of gaussians only. We delete the soft
  continuum component from model~A and refit the data. At low energies
  the fit now contains broad lines with a width on the order of
  150--180~eV, whereas the resolution of EPIC/MOS-2 has a width of
  $\sim 40$~eV at 1~keV. The absorption of the direct and scattered
  components is ``lower'' in eclipse than in egress (N$_{\rm H} \sim
  5.6 \times 10^{22}$ cm$^{-2}$ and N$_{\rm H} \sim 30 \times
  10^{22}$ cm$^{-2}$ in eclipse w.r.t N$_{\rm H} \sim 15.3 \times
  10^{22}$ cm$^{-2}$ and N$_{\rm H} \sim 82 \times 10^{22}$ cm$^{-2}$).

  We also investigated the soft component in the RGS dataset (see
  Fig.~\ref{low-flux_lines}). Due to the low RGS count rate it was not
  possible to set further constraints.

%
%

\subsection{Line Emission}
\label{line_emission}
  As in other HMXBs (e.g. \object{Vela~X$-$1} and \object{Cen~X$-$3};
  \citealt{sak99,ebi96}), \object{4U~1700$-$37} shows many emission lines in
  the eclipse and egress spectrum, when the continuum emission is at a
  minimum. Only some of these lines can be detected in the low-flux
  spectrum. We fit these lines as gaussians and list them in
  Table~\ref{linetable} if their significance is larger than
  2$\sigma$. For the lines detected in some, but not all spectra, we
  set an upper limit on the line flux in case of a non-detection.
  We do not correct for absorption in calculating the flux, since the
  geometry of the line forming region is not known.

  As the energy resolution of the EPIC/MOS-2 instrument is not
  sufficient to resolve all lines, many of them are blended. This
  complicates the line identification process (for which we use
  \citealt{joh85,dra88}), especially in the low energy range. The
  lines that we can identify are fluorescence emission lines from
  near-neutral species and discrete recombination lines from He- and
  H-like species.
  We will concentrate on the iron complex and the lines that are
  clearly resolved and report on their behaviour, moving from eclipse
  into egress and their presence in the low-flux spectrum.

\subsubsection{The Fe-line complex}
\label{ironcomplex}
  Figs.~\ref{egress} and~\ref{eclipse} clearly show that the Fe
  K$\alpha$ line around 6.4~keV is the strongest line in the eclipse
  and egress spectrum ($6.3 \times 10^{-4}$ and $1.5 \times 10^{-3}$
  photons cm$^{-2}$ s$^{-1}$, respectively). A quick look at the
  piled-up part of the lightcurve shows that the line is clearly
  detected throughout the orbit. However, in the low-flux spectrum the
  line has a strongly reduced flux ($5 \times 10^{-5}$ photons
  cm$^{-2}$ s$^{-1}$; Fig.~\ref{low-flux_lines}).
  In the eclipse and egress spectrum the measured line energy is
  consistent with an ionisation degree ranging from near-neutral iron
  to about \ion{Fe}{xvii}.

  A second, broad line is detected in the eclipse spectrum at a slightly
  higher energy around 6.5~keV. Its energy is consistent with Fe
  K$\alpha$ produced by more highly ionised iron (up to about
  \ion{Fe}{xxiv}). Since this line is very broad ($\sigma \sim
  0.3$~keV) and close to the 6.4~keV line, we consider a second
  possibility (as reported by \citealt{lab01} on {\it Beppo}SAX
  observations of \object{LMC~X$-$4}) in which two gaussians at 6.4~keV are used
  to mimic one line with a narrow core and broad wings. This does not
  result in a good fit. In egress we do not find evidence for a
  second, broad line around 6.5~keV, but a narrow line at 6.7~keV is
  present, which is consistent with Fe K$\alpha$ from almost
  completely ionised, He-like iron (\ion{Fe}{xxv}).
  In the low-flux spectrum we do not detect this line. With its energy and
  width fixed to the values found for the line in the egress spectrum,
  we can place an upper limit on its flux of $0.92 \times 10^{-5}$
  photons cm$^{-2}$ s$^{-1}$. This is much less than the measured
  flux in egress, similar to the strongly reduced flux of the Fe
  K$\alpha$ line around 6.4~keV.

  In addition to the Fe K$\alpha$ lines we detect the Fe K$\beta$ line at
  7.1~keV in the eclipse and egress spectrum. In the low-flux spectrum
  again only an upper limit can be placed on the flux of this line.
  Similar to the second K$\alpha$ line, the energy of the Fe K$\beta$
  line is slightly higher in egress than in eclipse ($7.11$ and
  $7.06$~keV, respectively). This is consistent with iron ionised up to
  about \ion{Fe}{xvii}, which is similar to the iron K$\alpha$ line at
  $6.4$~keV.

  The fluxes of the different iron lines show a similar behaviour. All
  the lines are much weaker in the low-flux spectrum than in the
  egress spectrum, and the fluxes of the near-neutral iron K$\alpha$ and
  K$\beta$ lines are higher in egress than in eclipse. However,
  there are some differences which become apparent when calculating the
  ratio Fe K$\beta$/K$\alpha$. In eclipse we find a value of
  K$\beta$/K$\alpha = 0.12_{-0.01}^{+0.01}$, which is in agreement with
  the value of 0.13 that is expected for an optically thin plasma
  \citep{kaa93}. However, in egress this ratio increases
  to K$\beta$/K$\alpha = 0.35_{-0.04}^{+0.04}$. Due to the high upper
  limit on the flux of the K$\beta$ line in the low-flux spectrum this
  ratio could be as high as 4.1.

  The presence of a K-edge from near-neutral iron at 7.117~keV alters
  the shape of the continuum around this energy. If the edge is not
  included in the model, this would result in an increased flux of the
  Fe K$\beta$ line as well as the value of the K$\beta$/K$\alpha$
  ratio. A deeper edge will thus decrease the K$\beta$/K$\alpha$
  ratio. An edge is clearly present in both the eclipse and the egress
  spectrum (see Figs.~\ref{egress} and~\ref{eclipse}), which is taken
  into account by the absorption component of the model. We do not
  find evidence for a deeper edge. If we put the abundance of Fe at
  zero the edge is not modelled by the absorption component. If we add
  an edge to the model separately, this results in an edge at slightly
  higher energy ($7.17_{-0.02}^{+0.01}$~keV). The flux of the K$\beta$
  line reduces to $\sim 3 \times 10^{-4}$ photons cm$^{-2}$
  s$^{-1}$. The K$\beta$/K$\alpha$ ratio then becomes $\sim 0.15$,
  which is consistent (within its errors) with 0.13 \citep{kaa93}.

\subsubsection{Line flux variation with orbital phase}
\begin{figure}[!t]
  \centering
  \resizebox{8.5cm}{!}{\includegraphics[angle=90]{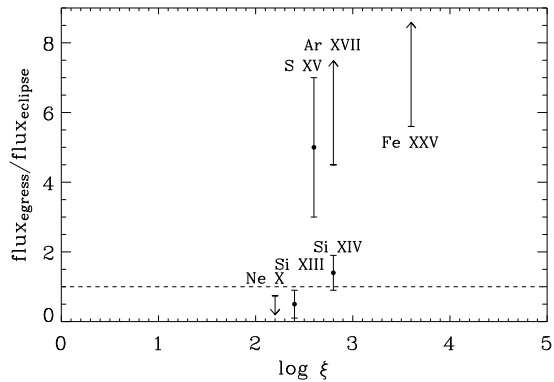}}
  \caption{The flux ratio of the lines in eclipse and egress as a
  function of the ionisation parameter ${\rm log~}\xi$. Lines
  corresponding to higher $\xi$ values become stronger going from
  eclipse to egress.}
  \label{line_ratios}
  \end{figure}

  We compare the flux from eclipse to egress of the different discrete
  recombination lines considering only the lines that we can clearly
  separate from the other lines and for which we have a unique
  identification. The fluxes of $\ion{Si}{xiv}~{\rm Ly\alpha}$,
  $\ion{Si}{xiii}~{\rm He\alpha}$ and $\ion{Ne}{x}~{\rm Ly\alpha}$
  remain constant within their errors, while the fluxes of
  $\ion{S}{xv}~{\rm He\alpha}$, $\ion{Ar}{xvii}~{\rm He\alpha}$ and
  $\ion{Fe}{xxv}~{\rm He\alpha}$ increase towards egress (see
  Fig.~\ref{line_ratios}).
  \begin{table}[!b]
  \caption[]{The value of the ionisation parameter $\xi$ corresponding
  to the reported emission lines.}
  \label{ionisation_par}
  \begin{flushleft}
  \begin{tabular}{p{0.8cm}p{0.6cm}|p{0.8cm}p{0.6cm}}
  \hline \hline
  \multicolumn{2}{c}{\rule[-2mm]{0mm}{6mm}H-like} 
  & \multicolumn{2}{|c}{He-like} \\ \hline
   \multicolumn{1}{l}{ion} 
  & \multicolumn{1}{l}{log $\xi$} 
  & \multicolumn{1}{|l}{ion} 
  & \multicolumn{1}{l}{log $\xi$} \\
  \hline 
  &&&\\
  \ion{O}{viii}   & $1.6$ & \ion{O}{vii}   & $ 1.0$ \\[0.05cm]
  \ion{Ne}{x}     & $2.2$ & \ion{Ne}{ix}   & $ 1.6$ \\[0.05cm]
  \ion{Mg}{xii}   & $2.6$ & \ion{Mg}{xi}   & $ 2.1$ \\[0.05cm]
  \ion{Si}{xiv}   & $2.8$ & \ion{Si}{xiii} & $ 2.4$ \\[0.05cm]
  \ion{S}{xvi}    & $3.2$ & \ion{S}{xv}    & $ 2.6$ \\[0.05cm]
  \ion{Ar}{xviii} & $3.4$ & \ion{Ar}{xvii} & $ 2.8$ \\[0.05cm]
  \ion{Fe}{xxvi}  & $4.2$ & \ion{Fe}{xxv}  & $ 3.6$ \\[0.05cm]
  \hline
  \end{tabular}
  \end{flushleft}
  \end{table}

  We use a power-law model with photon index $\alpha=1.4$ in the XSTAR
  photoionisation code \citep{bau01} to determine the ionisation
  parameter $\xi$ (see Sect.~\ref{discussion}) for the reported ions,
  assuming a photoionised plasma. The ionisation parameter $\xi$ is
  defined as
\begin{equation}
    \xi = \frac{L_{\mathrm X}}{n r^{2}},
\label{equation_xi}
\end{equation}
  where $L_{\mathrm X}$ is the bolometric X-ray luminosity,
  $n$ the number density at the distance $r$ from the
  ionising source in an optically thin plasma \citep{tar69}. The
  ionisation parameter then determines the local degree of ionisation
  of the plasma and can be used to predict the population of ions
  expected for a certain value of this parameter.
  
  The values of $\xi$ are listed in Table~\ref{ionisation_par}. We
  apply these values on the selection of lines and plot the ratio of
  the line fluxes in eclipse and egress as a function of $\xi$ (see
  Fig.~\ref{line_ratios}). We find that the lines with log $\xi
  \gtrsim 2.7$ are stronger in egress while the lines with a lower
  value for the ionisation parameter remain constant, as expected for
  a plasma that is photoionised by the X-rays from a central source.

%
%

\section{Discussion}
\label{discussion}
\begin{figure}[!t]
  \centering
  \resizebox{8cm}{!}{\includegraphics[]{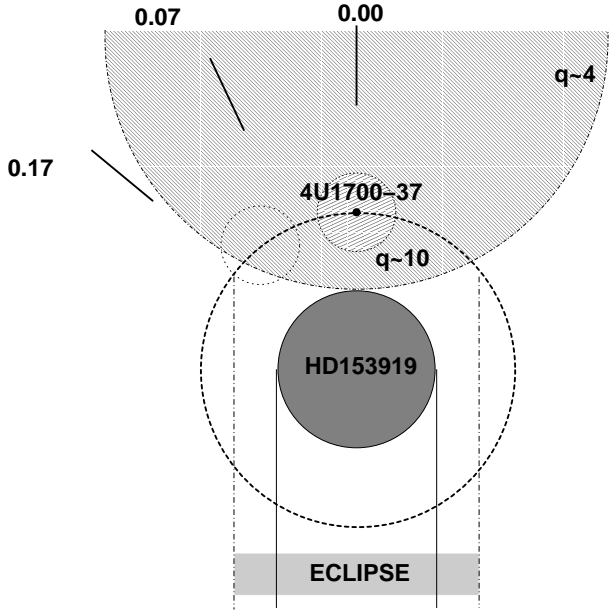}}
  \caption{This sketch of the geometry of 4U~1700$-$37 includes the size
    of the region where q $\lesssim 4$, the upper limit on the size of
    the ionised region reported by \citet{loo01}}
  \label{geometry}
  \end{figure}

  We interpret our observations including information obtained in
  other wavelength domains. The general picture we have in mind is
  that of a compact object, which accretes from the extended stellar
  wind of the O supergiant companion in which it is embedded. Due to
  this accretion an X-ray continuum is produced, which photoionises
  its near surroundings (see Fig.~\ref{geometry}). The degree of
  ionisation (expressed in terms of $\xi$) in this region decreases
  with distance from the X-ray source. Emission lines corresponding to
  higher $\xi$ values originate closer to the X-ray source
  (see. Eq.~\ref{equation_xi}).

\subsection{Continuum emission}
  We have shown that the high energy part of the continuum of
  \object{4U~1700$-$37} can in all cases be represented by a hard, direct
  component and a (Thomson) scattered component, which are modelled as
  two power laws with identical photon index, but different absorption
  coefficients. The amount of absorption on both components depends on
  the line of sight to the X-ray source and thus on the orbital
  phase.

  The direct and scattered component can be clearly separated in the
  egress spectrum, when the absorption of the direct component is
  high. The unabsorbed flux (F$_{\rm X}$) in the energy range
  0.3--11.5 keV of the direct component is $\sim 5 \times 10^{-9}$
  erg s$^{-1}$ cm$^{-2}$. Adopting a distance of 1.9~kpc
  \citep{ank01}, this yields an unabsorbed luminosity L$_{\rm X} \sim
  2.2 \times 10^{36}$ erg s$^{-1}$, consistent with previously
  reported values.

  During eclipse there is still evidence for the presence of the direct
  component, although its flux is strongly reduced (F$_{\rm X} \sim 1
  \times 10^{-10}$ erg s$^{-1}$ cm$^{-2}$, which gives L$_{\rm X} \sim
  5 \times 10^{34}$ erg s$^{-1}$). Note that our dataset does not
  cover a full eclipse, but starts at $\phi \sim 0.08$ according to
  the ephemeris of \citet{rub96}. \citet{rub96} derive an eclipse
  semi-angle $\theta_{\rm e} = 28\degr.6 \pm 2\degr.1$, using {\it
  BATSE} on {\it Compton} GRO, which has an energy range of
  20--120~keV. Therefore, the ``physical'' eclipse should end already
  at $\phi \sim 0.08$. However, at soft energies the eclipse lasts
  longer due to the strong absorption of soft X-rays by the stellar
  wind. \citet{hab89} found $\theta_{\rm e} = 44\degr.6 \pm 0\degr.6$
  in {\it EXOSAT} observations in the 2--10~keV energy range. This
  results in an ``end of the eclipse'' at $\phi \sim 0.12$, consistent
  with our observations. Due to this energy dependent definition of
  eclipse, a small contribution to the continuum with high absorption
  can still be expected.

  We modelled the low-flux spectrum at orbital phase $\phi \sim 0.25$
  in two ways, differing in how the high-energy part of the spectrum
  is fitted. In model~A (see Table~\ref{continua}) a highly absorbed
  direct component and a scattered component are used, with
  similar unabsorbed fluxes (F$_{\rm X} \sim 5 \times 10^{-10}$ erg
  s$^{-1}$ cm$^{-2}$; L$_{\rm X} \sim 2 \times 10^{35}$ erg
  s$^{-1}$). Model~B (see Table~\ref{continua}) contains only a direct
  component with an unabsorbed flux similar to that derived for
  model~A. This component is modelled as a power law modified by a
  high-energy cutoff and is the same model \citet{rey99} used to fit
  their {\it Beppo}SAX observations. For both models the flux of the
  direct component is a factor $\sim 9$ lower than in egress. In
  model~A the absorption on the direct component is relatively high
  (N$_{\rm H} \sim 2.0 \times 10^{23}$ cm$^{-2}$), whereas in model~B
  the absorption is lower (N$_{\rm H} \sim 7 \times 10^{22}$
  cm$^{-2}$). This value is consistent with calculations reported by
  \citet{hab89}, who predict N$_{\rm H} \sim 5-7 \times 10^{22}$
  cm$^{-2}$ at orbital phase $\phi \sim 0.25$.

  If we compare the flux of the direct component in egress to that in
  low flux, we find that a similar decrease in the scattered component
  should result in a non-detection. A strong reduction of the flux of
  the direct component and consequently of the scattered component as
  inferred from model~B can be explained by a reduction in accretion,
  such as the case in a ``clumped'' wind structure. \citet{sak99} also
  find evidence for clumping in the stellar wind of the optical
  companion of \object{Vela~X$-$1}. In case of Bondi-Hoyle accretion, often
  used to explain the production of X-rays in wind-accreting HMXBs
  \citep[e.g][]{nag89}, the amount of material accreted (and thus
  L$_{\rm X}$) scales as $\rho/v^{3}$, where $\rho$ is the density
  of the accreted material and $v$ its velocity. Therefore, the
  accretion is very sensitive to changes in the density- and
  (especially) in the velocity structure of the stellar wind
  \citep[see][]{kap93}. The X-ray lightcurve (see
  Fig.~\ref{lightcurve}) shows strong variability, which suggests a
  continuously changing accretion rate onto the compact object. Given
  the duration of the low-flux interval ($\sim 5$~ks) and assuming a
  typical wind velocity $v \sim 1000$~km~s$^{-1}$ the region should
  extend upto $\sim 5 \times 10^{11}$~cm (i.e. $\sim 0.4~{\rm
  R}_{\star}$; \citealt{cla02}).

  Model~A can only be consistent with this scenario if the scattered
  component is not affected by the drop in accretion, since this model
  predicts a scattered component with a similar unabsorbed flux as the
  direct component. This is only possible if it is produced
  sufficiently far out in the system, that it does not notice a change
  in accretion rate for $\sim 5$~ks. This would require a distance of
  $\sim 10^{14}$~cm ($\sim 70 {\rm R}_{\star}$), whereas the
  orbital separation of the compact object and the optical companion
  is $\sim 2~{\rm R}_{\star}$ \citep{hea92}. This is not consistent with
  the relatively high column density (N$_{\rm H} \sim 7 \times
  10^{22}$ cm$^{-2}$) toward the scattered component.

  A possible alternative scenario that can be described by model~A is
  the presence of an accretion wake, formed in front of the compact
  object, which creates a kind of eclipse that blockes the photons
  originating from the direct surroundings of the compact object. The
  presence of such a wake has been suggested by \citet{hab94}, based
  on a difference in the temperature of the bremsstrahlung component
  found in {\it Ginga} observations.
  In this scenario the interval of low flux should be locked to
  orbital phase, which is not consistent with other observations that
  report on low-flux intervals at different orbital
  phases. \citet{bor03} find a quiescence period in {\it Chandra}
  observations around $\phi \sim 0.68$. Therefore, model~A does not
  give a good physical description of the low-flux spectrum.

  In the case of an exponential cutoff, the values of $E_{c}$ and
  $E_{f}$ can not be determined very well, due to the energy range of
  the EPIC/MOS-2 camera, but using fixed values for $E_{c}$ and
  $E_{f}$ in eclipse and egress, model~B can be used to describe all
  three spectra. There is an indication for an increase of the photon
  index from eclipse and egress to the low-flux interval. Although the
  value of the photon index is correlated to the value of $E_{f}$, one
  can not exclude a change in the physical conditions of the accretion
  during the low-flux interval.

  We conclude that model~B probably best describes the physical
  conditions of the low-flux interval, in which the scattered
  component is not detected. A broader energy range is needed to
  further constrain the parameters of the high-energy cutoff.

\subsection{Soft excess}
  A soft excess is clearly present in all spectra and can be modelled
  with components of different physical nature, modified by an
  absorption which is consistent with the interstellar value for this
  source. The previously reported bremsstrahlung component
  \citep{hab92,hab94} results in a fit with a temperature that can not
  be constrained and a shape that mimics a power law. Also in the
  analysis of previously obtained data, the temperature of the
  bremsstrahlung component had to be fixed
  \citep[e.g.][]{hab94,rey99}. Therefore, we do not obtain more
  information on the physical origin of this component than with a
  simple power law model.

  Another possible model for the soft excess is blackbody
  emission. This component is often visible in HMXBs hosting an
  accretion disc (e.g. \object{LMC~X$-$4} and \object{Cen~X$-$3}; \citealt{lab01} and
  \citealt{bur00}, respectively). We can calculate the size of the
  emitting surface of the blackbody component:
  \begin{equation}
    R_{\rm km} = 3.05 \times 10^{9} d_{\rm kpc} F^{0.5} T_{\rm keV}^{-2}~,
  \end{equation}
  in which $R_{\rm km}$ is the radius of the emitting surface in km,
  $d_{\rm kpc}$ the distance to the source in kpc, $F$ the flux in erg
  s$^{-1}$ cm$^{-2}$ and $T_{\rm keV}$ the temperature in keV. Adopting a
  distance of 1.9~kpc \citep{ank01}, we find an upper limit for the
  radius of the emitting volume of 52~km for the low-flux
  spectrum. For the eclipse and the egress spectrum we find upper
  limits for the radius of 0.13 and 0.34~km, respectively. These latter
  values are very small. Adding the fact that the absorption is very
  low, the blackbody component can not originate from an accretion
  disc or (close to) the compact object. We therefore conclude that
  although the soft excess can be described by a blackbody component,
  a proper physical interpretation is lacking.

  The last possible continuum model which fits the soft excess is scattering. We
  examined both dust scattering and Thomson scattering. Dust
  scattering is energy dependent ($\propto 1/E^{2}$), which results in
  a steeper power law, mainly influencing the high-energy part of
  the spectrum. The difference between dust- and Thomson scattering
  can only be measured if the high energy part of the spectrum is not
  dominated by the strong direct component, i.e. during eclipse. Using dust
  scattering in eclipse does not improve the fit, whereas Thomson
  scattering does. Furthermore, this component has to originate from the
  outskirts of the system due to its low absorption. This is not
  consistent with a continuum that is absorbed in the extended stellar
  wind on its way out, before it is partly scattered. Therefore, we
  conclude that although Thomson scattering is a model that fits the
  data well in all cases, it is unlikely to be the physical origin of
  the soft excess.

  Recently, \citet{bor03} reported on {\it Chandra} observations of
  \object{4U~1700$-$37}. They claim that the soft excess can be explained by a blend
  of lines only. However, the statistics of their observations are
  low at these energies and they only report one unresolved line
  below 1~keV. We tried to fit the soft excess in our spectra in a
  similar way by removing the soft continuum component and using broad
  gaussians to create a pseudo-continuum.
  Although this results in stable fits, the lines are 4--5 times
  broader than the energy resolution of EPIC/MOS-2 at low
  energies. Furthermore, the absorption of the direct and scattered
  component is $\sim 3$ times lower in eclipse than in egress. We
  conclude that this model does not provide a satisfying physical
  description of the continuum. However, it does not exclude the
  possible presence of a blend of lines. Therefore, it remains a
  matter of debate whether the soft excess is real or due to a blend
  of unresolved emission lines.

\subsection{The Fe-line complex}
  The line spectrum consists of fluorescence emission lines of
  near-neutral species and discrete recombination lines of He- and
  H-like species. In eclipse and egress many lines are detected, while
  the low-flux spectrum shows only a few. Many lines are blended with
  each other or have multiple identifications, which complicates the
  analysis.

  One of the most striking results is that the iron K$\alpha$ complex
  around 6.4~keV is strongly reduced in flux in the low-flux spectrum
  (see Fig.~\ref{low-flux_lines}), while it is clearly visible in the
  rest of the (piled up) lightcurve. This is consistent with the
  reduced flux of the continuum above 6.4~keV and therefore another
  argument in favor of the scenario invoking a reduced accretion rate
  during the low-flux interval.

  Moving from eclipse into egress, we find an increase by a factor
  $\sim 2$ of the flux of the Fe K$\alpha$ line at 6.4~keV. This
  suggests that the formation region of the fluorescence lines from
  near-neutral elements is not far from the compact object. However
  the presence of discrete recombination lines from He- and H-like
  elements indicates a highly ionised region. In some systems an
  accretion disc may be the source of these emission lines
  (e.g. \object{LMC~X$-$4}; \citealt{lab01}). However, \object{4U~1700$-$37} is a wind
  accreting system, for which no evidence for an accretion disc has
  been found. Another possible scenario \citep[e.g. ][]{sak02} relates
  to the clumping of the stellar wind. If the wind is inhomogeneous,
  cool dense regions can form the near-neutral material, responsible
  for the presence of the fluorescence lines. This would result in a
  hybrid medium of cool dense and highly ionised regions.

  The expected K$\beta$/K$\alpha$ ratio for an optically thin plasma
  \citep{kaa93} should be around 0.13. In Sect.~\ref{ironcomplex} we
  show this is the case for eclipse, but in egress this is only true
  if the Fe edge is shifted to a slightly higher energy
  ($7.17_{-0.02}^{+0.01}$~keV). An edge at such energy can be produced
  by slightly more ionised iron, upto \ion{Fe}{iii}
  \citep{lot68}. Ionisation stages of iron like \ion{Fe}{iii} and
  \ion{Fe}{iv} are abundant in the stellar winds of OB supergiants and
  are important wind drivers. Although we can exclude the edge to be
  produced by \ion{Fe}{iv} or higher, this is not consistent with the
  results of \citet{vin00} who find that the stellar winds of OB-stars with
  temperatures higher than 25000 K are line driven by \ion{Fe}{iv},
  which recombines to \ion{Fe}{iii} at larger distances from the
  stellar photosphere, typically around ${\rm R} \sim 6-8~{\rm
  R_{\star}}$ \citep{vin99}. This is beyond the orbital separation of
  $\sim 2~{\rm R}_{\star}$ \citep{hea92} and thus outside the region
  probed by the compact object during the egress interval (see
  Fig.~\ref{geometry}).

\subsection{Recombination lines}
  We detect discrete recombination lines in both the eclipse and the
  egress spectra. The emission lines reported in
  Sect.~\ref{line_emission} can be identified with He- and H-like
  ions, with ionisation parameters in the range $2 \lesssim {\rm
  log~}\xi \lesssim 4$. If we compare the flux of the lines that are
  clearly separated and have a unique identification, as a function of
  ${\rm log~}\xi$, we see a systematic change in flux of some of the
  lines from eclipse to egress. The elements with ${\rm log~} \xi
  \gtrsim 2.7$, i.e. $\ion{S}{xv}~{\rm He\alpha}$,
  $\ion{Ar}{xvii}~{\rm He\alpha}$ and $\ion{Fe}{xxv}~{\rm He\alpha}$,
  grow in strength (see Fig.~\ref{line_ratios}), while
  $\ion{Si}{xiv}~{\rm Ly\alpha}$, $\ion{Si}{xiii}~{\rm He\alpha}$ and
  $\ion{Ne}{x}~{\rm Ly\alpha}$ remain constant within their
  errors. The observation of discrete recombination lines in an
  eclipse spectrum directly implies that the formation region extends
  beyond the size of the O supergiant. In our observations the eclipse
  interval covers a range in orbital phase of $0.07 \le \phi \le
  0.14$; therefore, the size of the ionised region should be $\sim
  0.5~{\rm R}_{\star}$ (see Fig.~\ref{geometry}). Because the lines
  with ${\rm log~} \xi \gtrsim 2.7$ are stronger in egress, the {\it
  Str\"omgren zone}, often defined as an ionised region with ${\rm
  log~} \xi \sim 3$, can not be too large.

  At these relatively high values of $\xi$ inside this ionised region
  ions like \ion{C}{iv}, \ion{Si}{iv} and \ion{N}{v} should be totally
  absent, resulting in an observable modulation of the UV
  resonance lines formed by these ions. This orbital modulation of UV
  resonance lines has been predicted for \object{4U~1700$-$37} by
  \citet{hat77}, but has never been observed \citep{dup78,kap93}. A
  possible explanation for the absence of the Hatchett-McCray (HM)
  effect in \object{4U~1700$-$37} is that the stellar wind has a non-monotonic
  (``turbulent'') velocity structure which can hide the presence of an
  ionisation zone from detection in an UV resonance line
  \citep{kap93}.

  \citet{loo01} derive the size of the ionisation zone for a few
  HMXBs, based on a wind model taking the non-monotonic velocity
  structure into account. The
  size of the ionisation zone can also be expressed \citep{hat77} in
  terms of the non-dimensional variable $q$, that is defined as
\begin{equation}
  q = \frac{\xi n_{\mathrm X} D^{2}}{L_{\mathrm X}},
\label{equation_q}
\end{equation}
  where $\xi$ is the ionization parameter, $L_{\mathrm X}$ the X-ray
  luminosity, $n_{X}$ the number density at the orbit of the ionising
  source and $D$ the distance from the center of the O supergiant to
  the ionising source. Substitution of eq.~\ref{equation_xi} in
  eq.~\ref{equation_q} then yields
\begin{equation}
  q = \frac{n_{\mathrm X} D^{2}}{n r^{2}}.
\label{equation_q_dimless}
\end{equation}
  In a medium with constant density, each value of $q$
  corresponds to a circle centered on the X-ray source, where the
  radius increases with decreasing value of $q$. In a stellar wind,
  the density falls off roughly as $r_{\star}^{-2}$, such that
  surfaces of constant $q$ are small and closed (high $q$ value; see
  Fig.~\ref{geometry}) or large and open with the ionising source
  off-center \citep[low $q$; ][]{loo01}. To each value of $q$
  corresponds a value of $\xi$, i.e. the higher the degree of
  ionisation, the higher the value of $q$. For \object{4U~1700$-$37}
  \citet{loo01} set an upper limit of $q \sim 4$ to the size of the
  ionisation zone based on the absence of the HM effect in \object{HD~153919}
  (see Fig.~\ref{geometry}). They predict that $q$ must be of the
  order 200, for ${\rm log~}\xi = 3$. Such a region would be much
  smaller than the size of the O supergiant. The detection of discrete
  recombination lines with high $\xi$ values during the eclipse
  spectrum indicates that the size of the ionisation zone must be
  larger, which is still consistent with the reported value for the
  upper limit, $q \lesssim 4$ \citep{loo01}.

%
%

\section{Summary and conclusions}
\label{conclusions}
  XMM-{\it Newton} observations of \object{4U1700$-$37} during eclipse
  ($0.07 \le \phi \le 0.14$) and egress ($0.14 \le \phi \le 0.17$), as
  well as a low-flux interval around $\phi \sim 0.25$, produced the
  following results:
  \begin{itemize}
  \item{The continuum is well represented by a combination of two
    power-law components: (i) a direct component produced by the X-ray
    source and (ii) a (Thomson) scattered component originating in an
    extended region surrounding the X-ray source. The extent of this
    region must exceed the size of the O supergiant as the scattered
    component is detected during eclipse.}
  \item{The low-flux interval is likely due to a lack of accretion
    onto the compact object due to a varying density and/or velocity
    structure of the accreting material, such as expected for a clumpy
    stellar wind of the optical companion.}
  \item{At low energies ($\lesssim 2$~keV) a soft excess is present,
    similar to what has been found in earlier observations with,
    e.g. EXOSAT and {\it Ginga}. We can exclude a thermal origin in
    the form of bremsstrahlung or blackbody emission.}
  \item{In all spectra we detect fluorescence line emission from
    near-neutral iron around 6.4~keV, probably produced by dense
    ``clumps'' in the stellar wind close to the compact object. In the
    low-flux interval this line is strongly reduced in flux in tandem
    with the continuum above 6.4~keV.}
  \item{The initially measured high Fe K$\beta$/K$\alpha$ ratio
    in egress can be explained by an Fe edge due to moderately ionised
    iron, upto \ion{Fe}{iii}. This is consistent with the ionisation
    degree of iron further out in the stellar wind of the optical
    companion, but not with the region probed by the compact object in
    egress.}
  \item{We detect many discrete recombination lines from He- and
    H-like species in both eclipse and egress. Some of the lines can
    be isolated, while most of them are blended with each other. The
    presence of these lines in eclipse directly implies that the
    formation region has a minimum size of around $0.5 {\rm
    R}_{\star}$. Some of the lines with a relatively high value of the
    ionisation parameter $\xi$ grow stronger towards egress. This
    implies that the ionisation zone can not be very large either.}
  \end{itemize}

%
%

\begin{acknowledgements}
  We would like to thank the VILSPA-team, Alex de Koter, Cor de Vries,
  Jelle Kaastra, Rolf Mewe, Ton Raassen, Richard Saxton, and Takayuki
  Tamura for the useful discussions and their help with the data
  reduction. We acknowledge support from NOVA.
\end{acknowledgements}

\end{document}